Impact of the antibiotic-cargo from MSNs on Gram-positive and

Gram-negative bacterial biofilms

Anna Aguilar-Colomer[a,b], Montserrat Colilla[a,b], Isabel Izquierdo-Barba[a,b],

Carla Jiménez-Jiménez[a,b], Ignacio Mahillo[c], Jaime Esteban[d*] and María

Vallet-Regí[a,b*]

[a] Dpto. Química en Ciencias Farmacéuticas, U.D Química Inorgánica y Bioinorgánica. Universidad Complutense de Madrid. Instituto de Investigación Sanitaria Hospital 12 de Octubre i+12. Plaza Ramón y Cajal s/n, 28040 Madrid, Spain.

[b] Centro de Investigación Biomédica en Red. CIBER-BBN, Madrid, Spain.

[c] Unidad de Bioestadística y Epidemiología. IIS-Fundación Jiménez Díaz. Av. De los Reyes Católicos, 2, 28040 Madrid, Spain

[d] Unidad de Microbiología Clínica. IIS-Fundación Jiménez Díaz. Av. De los Reyes Católicos, 2, 28040 Madrid, Spain

[*] Corresponding authors: E-mail address: jestebanmoreno@gmail.com (J. Esteban Moreno). Tel.: +34 91 394 1843; E-mail address: vallet@ucm.es (M. Vallet-Regí).

## Abstract

Mesoporous silica nanoparticles (MSNs) are promising drug nanocarriers for infection treatment. Many investigations have focused on evaluating the capacity of MSNs to encapsulate antibiotics and release them in a controlled fashion. However, little attention has been paid to determine the antibiotic doses released from these nanosystems that are effective against biofilm during the entire release time. Herein, we report a systematic and quantitative study of the direct effect of the antibiotic-cargo released from MSNs on Gram-positive and Gram-negative bacterial biofilms. Levofloxacin (LVX), gentamicin (GM) and rifampin (RIF) were separately loaded into pure-silica and amino-modified MSNs. This accounts for the versatility of these nanosystems since they were able to load and release different antibiotic molecules of diverse chemical nature. Biological activity curves of the released antibiotic were determined for both bacterial strains, which allowed to calculate the active doses that are effective against bacterial biofilms. Furthermore, in vitro biocompatibility assays on osteoblast-like cells were carried out at different periods of times. Albeit a slight decrease in cell viability was observed at the very initial stage, due to the initial burst antibiotic release, the biocompatibility of these nanosystems is evidenced since a recovery of cell viability was achieved after 72 h of assay. Biological activity curves for GM released from MSNs exhibited sustained patterns and antibiotic doses in the 2-6 µg/mL range up to 100 h, which were not enough to eradicate biofilm. In the case of LVX and RIF first-order kinetics featuring an initial burst effect followed by a sustained release above the MIC up to 96 h were observed. Such doses reduced by 99.9% bacterial biofilm and remained active up to 72 h with no emergence of bacterial resistance. This pioneering research opens up promising expectations in the design of personalized MSNs-based nanotherapies to treat chronic bone infection.

## 1. Introduction

Chronic bone infection is considered one of the most problematic biofilm-related infections [1-3]. In general, *Staphylococcus aureus* constitutes the most common bacterial genus associated with such infections [3-5]. Although, it has been reported that 5 and 25% of cases, are caused by different species of gram-negative bacteria, including a growing number of multidrug-resistant strains that hinders the treatment of these patients [6-8]. Recently, nanotechnology and nanocarrier-based approaches have irrupted into this landscape by bringing up innovative alternatives to prevent and combat infections [9,10]. Nanocarriers able to load, protect and locally deliver antimicrobial agents become ideal candidates to develop novel nanomedicines and reduce the side effects associated to conventional therapies [11-13]. Among them, Mesoporous Silica Nanoparticles (MSNs) exhibit unique structural, textural and chemical features and good biocompatibility [14-16], with a wide diversity of potential biomedical applications [17,18].

Numerous studies have proposed MSNs as antibiotic drug delivery nanosystems for antibacterial therapy [18-21]. Thus, MSNs have been engineered to exhibit bacteria/biofilm targeting capability [22-26], co-deliver multiple antibiotics [27], or carry on combinations of antibiofilm and bactericidal agents [28]. In general, these studies evaluate the antibiofilm efficacy of the whole nanosystem, studying the combined effect of different elements (targeting agents, multitherapy components etc.). However, little attention has been paid to investigate just the active role of the antibiotic-cargo loaded in mesostructure and their antibiofilm effect as function of the time. Furthermore, the majority of these studies are preliminary and qualitative, showing in most cases in vial release curves of antibiotic-cargo determined by different spectroscopy techniques.

Herein we report a systematic and quantitative study investigating the active doses released from different model nanosystems by determining the biological active curves and their direct antibiofilm effect on Gram-positive (*Staphylococcus aureus*) and Gram-negative (*Escherichia coli*) bacterial biofilms. Moreover, amine-functionalization of MSNs has been demonstrated as a suitable approach to provide MSNs themselves of bacteria/biofilm targeting capability relying on the affinity of positively-charged nanoparticles towards the negatively charged bacteria and biofilm [23]. On the other hand, amine-modified MSNs become ideal starting nanoplatforms for the subsequent grafting, via carbodiimide chemistry, of active targeting moieties for selective recognition of pathogenic bacteria.[25,29] For this reason, in the current research work pure silica and amine-modified MSNs were chosen as model nanoplatforms. Moreover, levofloxacin (LVX), gentamicin (GM) and rifampin (RIF), widely used for chronic bone infection treatment in clinical practice, were chosen and separately loaded into the different MSNs. The novelty of our research lies in the determination of the biological activity curves of the three antibiotics once released from MSNs at different time periods. To the best of our knowledge, previous research reports have focused on determining drug release kinetics without paying attention to test the antibiotic activity of the released drug concentrations. Moreover, the strength of this study is the systematic evaluation of the impact of the released antibiotic doses on preformed biofilms in terms of biofilm eradication and antimicrobial susceptibility. This pioneering research work is a good starting point towards the implementation of MSNs-based custom-made advanced therapies designed to meet individual patient needs.

## 2. Experimental

## 2.1 Synthesis and functionalization of MSNs

MSNs were synthesized following the modified Stöber method by using hexadecyltrimethylammonium bromide (CTAB, Sigma-Aldrich) as structure directing agent [30]. The functionalization of MSNs with amino ($–NH_2$) groups was carried out by post-synthesis grafting of (3-aminopropyl)triethoxysilane (APTES, 99%, ABCR) under anhydrous conditions and inert atmosphere [31,32]. Amine-modification of MSNs plays a dual role: firstly, lead to positively-charged nanocarriers as bacteria/biofilm-targeted models [24,25]; secondly, it can be considered a pre-conditioning stage for the subsequent anchorage of biologically active molecules [17]. The details are shown in the Supporting Information.

## 2.2 Antibiotics loading

Levofloxacin (LVX, ≥98.0%), gentamicin sulfate (GM) and rifampin (RIF, ≥97.0%) were purchased from Sigma-Aldrich and used as received. These antibiotics have been chosen based on their broad spectrum activity against Gram negative and Gram positive bacteria and as they are commonly used, alone and in combination, in many types of infections [33]. **Figure 1** displays the structure and main chemical properties of the three chosen antibiotics. The antibiotic loading into MSNs was carried following the impregnation method reported by Cicuéndez *et al* [34]. The total amount of antibiotics loaded into MSNs was determined from the carbon content of samples quantified by elemental CHN chemical analyses. The details are shown in the Supporting Information.

## 2.3 Characterization of the nanosystems

All MSNs were submitted to a deep physicochemical characterization by using X-ray diffraction (XRD), transmission electron microscopy (TEM), $N_2$ adsorption, dynamic light scattering (DLS) and zeta ($\zeta$)-potential, elemental chemical analysis and Fourier transform infrared spectroscopy (FTIR). The details are shown in the Supporting Information.

## 2.4 In vitro stability test of MSNs matrices

With the aim of evaluating the stability of MSNs in the delivery medium, *in vitro* degradation tests were performed. The experimental conditions were similar to that of the release experiment (see section 2.5) but replacing TSB by phosphate buffer saline (PBS 1x, Sigma-Aldrich). After different incubation times, the stability of samples in terms of morphology and structural order was monitored by TEM. Unloaded pristine MSN served as control during the experiments.

## 2.5 *"In vial"* release assays

To determine the release curves of LVX, GM and RIF from MSNs, 10 mg of antibiotic-loaded MSNs were suspended in 0.5 mL of PBS 1x. Then, 680 µL of this suspension were placed on the upper chamber of a Transwell permeable support with a 0.4 µm polycarbonate membrane (Transwell® 6-well plate, Corning, USA). The well was filled by adding 3.1 mL of PBS to the lower chamber and the plate was incubated at 37 ºC under orbital stirring at 100 rpm in the absence of light. At selected time intervals of 0.5, 2, 4, 6, 24, 30, 48, 72 and 96 h, the PBS was collected from the lower chamber and replaced by 3.1 mL of fresh PBS. The release products were frozen at -80 ºC until later assays. The design of the "*in vial*" cargo release experiments to determine the LVX, GM and RIF active doses from MSNs was similar to that of in vitro experiments but replacing PBS by TSB + 1% glucose medium.

The amount of LVX released from MSNs was determined by fluorescence spectroscopy with ⋏ex = 292 nm and ⋏em = 494 nm in a BioTek Spectrofluorimeter (BioTek Instruments GmbH, Germany). First, a calibration curve was established in a concentration range from 0.01 to 12 µg/mL.

The amount of GM released from MSNs was monitored by high-performance liquid chromatography (HPLC) in an Alliance automatic analysis coupled to a Model#2696 photo-diode array detector and controlled by the software Empower2 (Waters, Massachusetts, US). For the assay, a 150 x 4.6 mm pre-packed analytical Mediterranea™ Sea18 column (Teknokroma, Spain) containing 5 µm C18 functionalized silica beads was used. The isocratic mobile phase was composed by acetonitrile and water (80:20, v/v) delivered at 1.0 mL/min at 25ºC. The injection volume was 10 µL. GM was detected by ultraviolet spectroscopy at 211 nm with a total chromatogram time of 10 minutes [19]. A calibration curve using standard concentrations of 1, 1.2, 1.4, 1.6, 1.8 and 2 mg/mL was obtained.

The amount of RIF released from RIF-loaded MSNs was monitored by ultraviolet spectroscopy at 474 nm in a BioTek Spectrofluorimeter (BioTek Instruments GmbH, Germany). The calibration curve was obtained by using standards with concentrations in the 0.88 to 28.1 µg/mL range.

## 2.6 Microbiological assays

To evaluate the antimicrobial capacity of the MSN-based nanosystems three separate experiments were carried out: the first one consisted on determining the biological activity curves of the antibiotic doses released from MSNs at given time periods; the second one aimed at evaluating the capability of such doses to eradicate preformed bacterial biofilms. Lastly, the third one focused on detecting the possible appearance of bacterial resistance. However, since the current manuscript focuses on evaluating determining the biological activity of the antibiotic cargo released from the different matrices, an in vitro experiment was designed by following an adaptation of a previously reported method [35]. The details are shown in the Supporting Information. Both *E. coli* ATCC 25922 and *S. aureus* ATCC 29213 strains were used. Due to the broad spectrum of action of LVX, it was tested on both bacteria. However, GM and RIF was tested on *E. coli* and *S. aureus*, respectively, because they are more specific.

**Biological activity curves**: Determination of the active antibiotic-doses released from MSNs was performed by *disc diffusion test* (see S.I.) for each bacteria and antibiotic as previously described Aguilera-Correa *et al* [36]. To obtain a standard curve, discs were also impregnated with a known concentration of each commercial antibiotic. Then, the corresponding biological activity of each dose released from different matrices was calculated. Moreover, biological activity curves were adjusted to the appropriate kinetic model, as it will be discussed below, which allowed to determine the principal fitting parameters, *i.e.* kinetics constant a maximum antibiotic active concentration.

**Antibiofilm activity**: The direct antimicrobial effect of antibiotic-cargo released from MSNs matrices on biofilms was performed following the recommendations of Stepanovic *et al* [37] for the preparation of mature biofilms for each bacterial strain. After obtaining 48 h biofilms from *E. coli* and *S. aureus* ($1\times10^8$ CFU/mL were added to each well), the extracts released from MSNs materials at different times were added. The plate was incubated 24 h at 37ºC-5% $CO_2$ and after that time, the biofilms were washed and sonicated (Thermo Fisher Scientific, USA) for 5 minutes to disrupt the biofilm. Serial dilutions were made and seeded using the *dropplate* technique in TSS medium plates (BioMérieux, France) to determine the number of CFU/mL for each tested release time [38]. The plates were incubated 24 h at 37ºC-5% $CO_2$. Negative control well contained only sterile broth. Positive control contained a known concentration to determine if the effect of the antibiotic is only produced by itself or if there is some kind of synergy with the releasing process. This experiment was performed six times for each antibiotic/strain.

**Preliminary study of the direct effect of MSNs on bacterial biofilms:** As an illustrative example, a preliminary study showing the direct effect of MSN-LVX and MSN-NH$_2$-LVX nanosystems on *E. coli* biofilm has been performed. As negative control antibiotic-free samples (MSN and MSN-NH$_2$) at a concentration of 75 µg/mL was used. On the other hand, free LVX at the maximum concentration released from the nanosystems ($\approx$ 4 µg/mL) was employed as positive control. These experiments have been carried out at different incubation times (2 and 24 h) to determine the advantage, in terms of antimicrobial efficacy, of using these nanoparticles when they act as drug reservoirs at local level. Briefly, after preparing different plates of 48 h *E. coli* biofilms ($1x10^4$ CFU/mL of bacteria added), the nanosystems were added at a 75 µg/mL concentration and then incubated during different times at 37º C in an oven. After the end of time, the biofilms were washed and sonicated for 5 min. Serial dilutions were prepared for each sample and they were seeded in TSA medium plates using the drop plate technique. The determination of CFU/mL was carried out after incubating the seeded plates for 24 h at 37 ºC. Biofilm controls only containing sterile culture medium have been carried out. All experiments have been performed in three different experiments with two replicas in each experiment.

**Bacteria Susceptibility test**: Bacterial susceptibility testing was performed using *E-test strips* (BioMérieux, France) [39]. Briefly, the different active doses from different matrices were treated during 24 h onto preformed *E. coli* and *S. aureus* biofilms grown 48 h. Subsequently, the biofilms were washed with PBS 1x and disrupted. The resulting bacterial rest (planktonic bacteria) were incubated for 4 h at 37ºC in an oven at 5% of $CO_2$ atmosphere. Then, they were inoculated on a plate of Müeller Hinton medium (BioMérieux, France) and an E-test strip impregnated with an increased antibiotic concentration was placed on the plate. After, all plates were incubated during 24 h at 37 ºC - 5% $CO_2$. The minimum inhibitory concentration (MIC) value before and after treatment were registered and compared to detect the appearance of resistance.

**Statistical analysis**: Data were described by median and ±S.D. Due to the small sample size of groups (n=6), statistical comparisons were performed with non-parametric methods. Comparisons between control and different times were made using Wilcoxon rank-sum test. For all comparisons the level of significance was set at 0.05. Statistical analyses were performed using R program 3.6.0 (R Foundation for Statistical Computing, Vienna, Austria).

## 2.7 Cell Biocompatibility assays

Since antibiotic-loaded MSNs are envisioned for bone infection treatment, it is necessary to study their effect on bone cells. To this aim, a cell proliferation experiment was carried out using mouse osteoblasts, specifically MC3T3-E1 cell line. MC3T3-E1 cells were seeded in 96-well plates (5000 cell/well) and treated with MSNs concentrations of 10, 25, 50 and 75 µg/mL. After 2 h of treatment, the cells were washed three times with PBS, and they were cultured in medium until 24h or 96 h. After such time periods, cell proliferation was analyzed using Thiazolyl Blue Tetrazolium Bromide (M5655, Sigma-Aldrich) following the manufacturer's instructions. During the assay, XTT tetrazolium is reduced until the blue crystals formation by mitochondrial dehydrogenases of living cells, which was detected colorimetrically at 570 nm using (Sinergy 4, BioTek, USA). For all the experiments, cell proliferation was expressed as the percentage of untreated cells. *In vitro* data were expressed as mean ±S.D. Further, we used Student's tests to determinate the significance. P < 0.05 was considered significant.

## 3. Results and discussion

### 3.1 Characterization of nanocarriers

The morphology and structural characterization of the resulting nanocarriers were performed by XRD (see **Figure S1**) and TEM studies. Both techniques confirm the ordered mesoporous structure of these nanosystems, showing a 2D-hexagonal arrangement (*p6mm* plane group) similar to a honeycomb-like structure (MCM-41 type). Moreover, TEM images show a *quasi-spherical* morphology with a particle diameter size around 150-200 nm. The monitoring of the nanosystems after functionalization and loading process by TEM show that neither the surface modification nor the antibiotic loading affects either the morphology or the structural properties, which remained constant, demonstrating their robustness after loading and functionalization processes. Notice no significant variation in the cell parameter ($a_0$) before and after functionalization and loading (see Table S1 of SI). Since these matrices will act as nanocarriers of different antibiotics, which will be gradually released, it is important to know their *in vitro* stability. As expected these nanosystems suffer degradation as it can be observed in TEM images, which reveal that the semispherical morphology is preserved, showing a loss of the internal mesopore structure after long incubation time. These results are in good agreement with those reported by Paris *et al*, [40] which demonstrated the total mesostructure loss after 9-12 days of incubation time. In fact, this loss is due to the silica lixiviation into free silicon species in the form of harmless monosilicic acid by-products that, after entering the bloodstream or lymph, would be well-excreted by the urine. As a representative example, **Figure 2** displays TEM images corresponding to pristine unloaded MSN before and after 9 days in PBS 1x solution at physiological conditions. To obtain more detailed information about MSNs after a period of 9 days see **Figure S2**.

To investigate the colloidal stability of nanoparticles in physiological media an additional experiment was carried out by incubating MSNs in cell culture medium (Dulbecco's Modified Eagle's Medium, DMEM, supplemented with 10 % Fetal Calf Serum, FCS) for 48 h and recording the hydrodynamic diameter ($D_H$) by DLS (see SI for further details). In addition, $\zeta$-potential measurements of nanoparticles in different media, namely water, PBS and DMEM-10% FCS, were performed. The results are collected in **Figure 3**. It was observed that, compared to the results obtained when MSNs were incubated in PBS (0% FCS), the $D_H$ of MSN and MSN-NH$_2$ underwent a 41% and 21% increase after 48 h in in culture medium (10% FCS). This is consistent with the well-known trend of this type of nanocarriers to aggregate and experience nonspecific protein adsorption (formation of a "protein corona") when encounter the complex biological milieu.[41,42] Moreover, upon exposure to 10% FCS, the $\zeta$-potential of MSN became substantially less negative, which is consistent with the formation of a protein corona. On the contrary, the $\zeta$-potential of MSN-NH$_2$ turned more negative after incubation with serum, in good agreement with previously reported data for positively charged nanoparticles.[43,44] Since this protein corona will define the biological identity of MSNs and their interaction with cells,[43,44] it may negatively affect their efficiency and final fate in vivo. Thus, when envisioning clinical application of these drug delivery nanoplatforms, tailoring the physical-chemical surface properties of MSNs by grafting of polyethyleneglycol (PEG),[45] or zwitterionization,[46] could contribute to reduce the non-specific proteins adsorption.

$N_2$ adsorption porosimetry measurements provided information about the textural properties of MSNs before and after being loaded with the antibiotics. The adsorption-desorption $N_2$ isotherms are in all cases typical of MCM-41 type mesoporous materials with parallel cylindrical pores (see **Figure S1**) [47]. The appropriate treatment of data allowed determining the main textural parameters, namely surface area ($S_{BET}$), pore volume ($V_P$) and pore diameter ($D_P$) (**Table 1**). It is observed that all these parameters experience a decrease in MSN-NH$_2$ compared to pristine MSN, which accounts for the fruitful post-grafting of organic moieties to the silica surface [48]. In this regard, with the aim of preferentially grafting the aminopropyl moieties to the external surface of MSN, the functionalization process was accomplished by the post-synthesis method with the surfactant molecules still present inside the pores of MSN, as described in the experimental section. Thus, the observed decrease in the $V_P$ and $D_P$ values of MSN-NH$_2$ compared to those of MSN could be attributed to the partial blockage of the pore entrances, as reported elsewhere.[22,23]

After being loaded with GM or RIF both MSNs showed a reduction in $V_P$ and $D_P$, pointing to the efficient confinement of the antibiotic molecules into the mesoporous cavities. This effect was more noticeable in the case of MSN-NH$_2$, where $V_P$ experienced a decrease of 94% and 65% after GM and RIF loading, respectively. On the contrary, after LVX loading, neither $V_P$ nor $D_P$ of both MSNs underwent significant variations. With the aim of understanding all these findings, which could be related to the course of the loading process, the amount of antibiotic incorporated to the different MSNs was determined by elemental chemical analysis (**Figure 4** and **Table S2**). Different loading performances were observed depending on the surface chemistry of the host mesoporous matrix and the chemical nature of the guest antibiotic molecule (**Figure 1**). Thus, in the case of pure silica MSN, the GM and RIF loadings were greater than LVX loading, in good agreement with the higher decrease in the textural properties observed for MSN-GM and MSN-RIF. The attractive electrostatic interactions taking place under the aqueous loading conditions between protonated amino groups of GM molecule and deprotonated silanol groups (–Si-O$^-$) of the silica surface would account for the high drug loading found in MSN-GM. When comparing RIF and LVX loadings in MSN, the amphoteric nature of both antibiotics should be considered (**Figure 1**). Similar interactions would take place between basic amine group of piperazine group present in both RIF and LVX molecules and silanol groups in the surface of MSN under the anhydrous loading conditions. However, the stronger acidity of hydroxyl group (–OH) of RIF (pKa 1.7) compared to the acidity of carboxylic acid group (–COOH) of LVX (pK$_a$ 5.59) (**Figure 1**) would promote 1stronger hydrogen bond interactions with the –SiOH of MSN (predominating under the anhydrous loading conditions). This fact would support the higher drug loading in the case of MSN-RIF compared to that of MSN-LVX. Regarding MSN-NH$_2$, the highest antibiotic loadings were found for RIF and LVX, owing the attractive interactions between amino groups of functionalized matrix and acid groups present in these two antibiotics. Again, the stronger acidic character of –OH group in RIF compared to the acidity of –COOH in LVX (**Figure 1**) would produce the highest antibiotic loadings in MSN-NH$_2$-RIF. Finally, when comparing GM and RIF loadings in MSN and MSN-NH$_2$ it is observed that the antibiotic loadings were greater in pristine MSN, which can be explained on the basis of the higher affinity of amino groups of these aminoglycoside antibiotics towards the silanol groups (Si–OH) of non-functionalized nanoparticles. However, this fact, at first glance, would not explain the abrupt decrease of the textural properties observed in MSN-NH$_2$ compared to pristine MSN after GM and RIF loading.

These results can be understood by taking into account the followed functionalization route itself, post-synthesis, which leads to a predominant attachment of the organic groups to the external surface of mesopores [48]. Thus, aminoglycoside molecules would penetrate into the mesoporous cavities to interact via H-bonding with the unreacted Si–OH groups mainly present in the inner pore surface of MSN-NH$_2$. On the contrary, LVX is a fluoroquinolone that exhibits hydrophobic and acid character [49], tending to interact with basic aminopropyl groups mostly present in the outer mesopores surface of MSN-NH$_2$. This would agree with the preservation of the textural parameters, mainly V$_P$ and D$_P$, after MSN-NH$_2$ is submitted to the LVX loading procedure. In view of these results, it can be concluded that the different host-guest interactions are the driving forces that govern antibiotics loading into MSNs and they are also foreseen to play a key role in drug release profiles and availability.

## 3.2 Determination of active antibiotic doses from MSNs

The active concentrations of the different antibiotics after release from MSNs at the different tested times against *E. coli* and *S. aureus* bacteria are displayed in **Figures 5** and **6**, respectively. For comparative purposes antibiotic doses released from MSNs during "*in vial*" experiments are also displayed in graphs. It should be noticed that in all cases the plot of released and active concentrations versus time provides similar patterns, which indicates that the antimicrobial activity of the antibiotic is proportional to the amount of drug released from MSNs. Finally, the released antibiotic concentrations are *ca.* 10-fold higher than the active concentrations, pointing to a partial loosening of the antimicrobial activity of the drug during the different stages it undergoes, from the loading process to the determination of active doses. Nonetheless, despite of this partial loss of antibiotic activity its antibacterial efficacy is positive enough, which is also evidenced during the antibiofilm activity assays, at it will be discussed in section 3.3.

With the aim of understanding the evolution of antibiotic active concentrations from MSNs over time, experimental data were fitted to a first order exponential decay model (Eq. 1):

$$C_t^A = \text{B e}^{-\text{mt}} \qquad (1)$$

where $C_t^A$ is the concentration of active drug after $t$ time of release, $B$ is the maximum drug concentration, and $m$ is the decay kinetic constant. The resulting parameters derived from the fit of experimental data to Eq. 1 are displayed as insets in **Figures 5** and **6**.

Studies of active concentrations of the LVX released from MSN against *E. coli* revealed an initial maximum active concentration of 34.04 μg/mL after 0.5 h of assay. This agrees with the initial rapid LVX release or burst effect observed during the "in vial" antibiotic release assays from MSNs (see SI, Figure S3). Subsequently, a progressive drop of the LVX active concentration *vs.* time was observed, reaching values of 0.45 μg/mL after 96 h of assay. Such behavior supports the release profiles obtained during the *"in vial"* experiments, where the initial burst effect is followed by a more sustained antibiotic release in both tested MSNs. Nonetheless, the decrease in the LVX active concentration is more pronounced in the case of the LVX released from MSN-NH$_2$, with a LVX active concentration of 48.9 μg/mL detected in the first half hour, in agreement with the higher antibiotic loading of functionalized sample compared to pristine MSN (**Figure 4**). On the other hand, the decrease in active concentration of LVX over time was more pronounced by detecting <0.1 μg/mL after 48 h, which corresponds to the test detection limit for this antibiotic and bacteria. All these findings are in good agreement with the results derived from the fit of the experimental data to Eq. 1, which provided exponential decay constant values, $m$, of 0.06 h$^{-1}$ and 0.24 h$^{-1}$ for LVX released from MSN-LVX and MSN-NH$_2$-LVX, respectively. The 4-fold faster decay in the active concentration of LVX released from MSN-NH$_2$ could be attributed to the predominant location of the antibiotic molecules in the external surface of the mesopores in funtionalized nanoparticles, as above discussed.

This behavior was not shared by the GM released from MSN-GM and MSN-NH$_2$-GM. GM exhibited sustained decay profiles, with active concentrations against *E. coli* in the 6.3 to 1.7 µg/mL range in both systems. In addition, this oscillation in active concentrations was detected in each of the repetitions, complicating the reproducibility of the results. For comparative purposes experimental data were also fit to Eq. 1. Although poor goodness of the fit was obtained, the resulting low $m$ values accounts for relatively slow GM release from both MSN and MSN-NH$_2$ materials, in good agreement with the results derived from "in vial" release experiments (See SI. **Figure S4**). This slow diffusion-driven release would be explained by the strong attracting interactions at pH 7.4 between protonated amino groups (–NH$_3^+$) of GM and deprotonated silanol groups (–SiO$^-$) prevailing in the inner mesoporous channels (See SI, **Figure S6**).

Biological activity tests against *S. aureus* of LVX and RIF released from MSNs provided fast-decay kinetic profiles similar for both antibiotics and materials (**Figure 6**). The predominating *zwitterionic* nature of both antibiotics at physiological pH would account for such parallelism (See SI. **Figure S6**). However, the LVX active concentration values were *ca.* 10-fold greater than those of RIF, being the maximum LVX active concentration 38.09 and 42.94 µg/mL for MSN-LVX and MSN-NH$_2$-LVX, respectively. The LVX active concentration was similar in *S. aureus* and *E. coli*, showing in both cases a more pronounced fast-decay in MSN-NH$_2$-LVX than in MSN-LVX. The minimum LVX active concentration detected was 0.82 µg/mL, which corresponds to the detection limit of the test for this antibiotic. Although *S. aureus* is more sensitive to RIF than to LVX and the amount of loaded drug was higher for the former, the active concentration released was lower (see SI, **Figure S5**). This could be explained by the low solubility of RIF in physiological media,[50] which, in fact, is one of the key parameters affecting bioavailability.[51] On the other hand, during the first hours the active concentrations of RIF released from MSN were higher than those released from MSN-NH$_2$, which agrees with the 2-fold higher amount of loaded antibiotic quantified for the former material. Thus, the maximum active concentration of RIF found was 4.12 μg/mL for MSN-RIF and 2.01 μg/mL for MSN-NH$_2$-RIF.

There were no significant differences in the decrease of the active RIF concentration released *versus* time, which agrees with the similar *m* values, 0.06 h$^{-1}$ and 0.09 h$^{-1}$ for MSN-RIF and MSN-NH$_2$-RIF, respectively. This fact would support the predominant location of the antibiotic into the mesoporous cavities in both materials, as above discussed. Due to the high sensitivity of *S. aureus* with the RIF the minimum active concentration detected was 0.01 μg/mL for MSN-RIF and 0.05 μg/mL for MSN-NH$_2$-RIF. Both antibiotics were easily reproducible.

### 3.3 Antibiofilm activity

Once the curves of active antibiotic doses and their kinetic fittings were determined, the impact onto *E. coli* and *S. aureus* biofilms was studied. Concerning *E. coli* biofilms, all active concentrations of LVX released from both MSNs showed a statistically significant biofilm reduction ($p < 0.05$) (**Figure 7**). There was no relationship between the active concentration released and the effect on the biofilm, since the biofilm was notably reduced between 99.8% and 89.2 at all tested times. On the contrary, the slow and sustained GM release did not provoke significant biofilm reduction, since the released antibiotic doses in the 2-6 µg/mL range up were not enough to eradicate it, in agreement with previous results [52].

In the case of *S. aureus*, in general, all the LVX doses released from MSNs notably reduced the biofilm, even though there were differences in the duration of the antibiofilm activity depending on the host matrix. Hence, the progressive decrease of the active LVX concentrations released from MSN-LVX allowed to maintain such effect over time, reducing the biofilm by 76.7% at 96 h and achieving a 99.4% reduction in the first hour (**Figure 8**). On the other hand, LVX released from MSN-NH$_2$-LVX showed an absence of antibiofilm effect at 96 h. The rest of the active concentrations released from this matrix significantly reduced the biofilm (p<0.05), reaching a 99.7% reduction at 2 h. Such dissimilarity can be attributed to the difference in the release kinetic constant between both nanosystems, being the LVX release faster from MSN-NH$_2$-LVX that from pristine MSN, as seen before (**Figure 6** and **Figure S3**).

The low active RIF concentrations released from MSNs, managed to significantly reduce the biofilm of *S. aureus* at all times. At 96 h, an active concentration of 0.012 µg/mL reduced the biofilm by 50%. Although the amount of released RIF was half from MSN-NH$_2$, the concentration released was able to reduce *S. aureus* biofilm up to 99% at 0.5 h and keeping such activity until 72 h of releasing. However, biofilm reduction was not as effective after 96 h of testing for both matrices (**Figure 8**).

### 3.4 Direct effect of antibiotic loaded MSNs onto bacterial biofilms

To evaluate the advantage of using the MSNs as antibiotic reservoir and/or dosing nanosystems at local level, a preliminary study with LVX-loaded MSNs (MSN-LVX and MSN-NH$_2$-LVX) onto *E. coli* biofilms was carried out. As positive control, the maximum amount of LVX released by the nanosystems was used. The results are collected in **Figure**

**9**. Both LVX-free MSNs produced a slight antimicrobial effect on *E. coli* biofilm, although this effect is more pronounced in the case of MSN-NH$_2$ sample. These results are in agreement with previous results with amine functionalized MSNs, which produced greater interaction with the bacteria wall and/or biofilm surface and triggered a certain effect on the biofilm. On the contrary, in the case of LVX-loaded MSNs, there is a remarkable reduction of biofilm that is comparable to the effect of isodose free LVX after 2 h of incubation. This reduction is maintained after 24 h, though the antimicrobial effect is more pronounced in the case of LVX-loaded MSNs compared to solely LVX. More in-depth studies using different antibiotics and different strains as a function of incubation time (including long-time periods) are being carried out and will be the subject of further studies.

### 3.5 Bacteria Susceptibility test

The emergence of resistant mutant bacteria when an infection is treated with monotherapy is a common event [39]. In fact, the ease of *S. aureus* in becoming resistant when it is treated with RIF is well-known [53]. Although the emergence of *E. coli* LVX- and GM-resistant mutants is infrequent, sensitivity studies of *E. coli* against LVX and GM and *S. aureus* against LVX and RIF from MSNs were necessary to detect possible changes in the sensitivity of these bacteria. The results showed that *E. coli* from biofilm maintained sensitivity to LVX and GM when it was treated during 24 h with the antibiotic cargo released at 2, 24 and 96 h from MSNs. The same behavior was noticed in *S. aureus* when it was treated 24 h with the LVX/RIF cargo released at the same time periods. MIC values are shown in SI. This result contrasts with the study performed by Aguilar-Colomer *et al* [54], which reported the appearance of *S. aureus* resistant mutants after treatment with the RIF released from SBA-15 matrix. According to the results of this work it is determined that the amount of antibiotic released from MSN was able to eliminate the greatest number of bacteria, reducing the probability of the emergence of resistant mutants. However, other variables such as inoculum size, exposition time, and the potential intrinsic antibacterial activity of the nanoparticles, could have also a role in avoiding the appearance of resistant strains. Regarding the application of these nanoparticles in human body, they are envisioned to be locally administered via intra-osseous injection of a nanoparticle suspension at the bone infection site. This local administration is an advantage, because the toxicity issues and collateral damages associated to the conventional systemic administration are minimized. [55]

Once there, nanoparticles are expected to reach the biofilm by passive targeting relying on the enhanced permeability and retention (EPR) effect, which has been evidenced that is not only present in tumors but also in biofilm-derived infections.[56,57] In this regard, the possibility to tailor the physical-chemical surface properties of MSNs is an added value, since different active targeting moieties can be attached to the outermost surface of nanoparticles to target biofilm and/or bacteria and therefore increase the efficiency and/or complement passive targeting. [22,23,58,59] Nonetheless, it is necessary that once internalized, the antibiotic is released in an efficient fashion. Ideally, the nanocarrier should release a high antibiotic concentration at the beginning of the treatment and subsequently release lower antibiotic concentrations for a prolonged time period from days to even weeks. In this sense MSNs synthetized in the current research work fulfills such ideal drug delivery performance for LVX and RIF antibiotics, which is characterized by an initial burst release followed by a sustained release where drug doses are lower but still effective. Thus, this two-steps release profile would prevent bacterial growth and adaptation to the treatment.

### 3.6 Cell Biocompatibility

Finally, *in vitro* preliminary assays studying the possible citotoxicity of all nanosystems at different concentrations (10-75 μg/ml) were performed. At 24 h, cytotoxicity studies with MC3T3-E1 cells showed that pristine MSN did not present toxicity even at the highest concentration of nanoparticles studied. However, MSN-NH$_2$ displayed a small cell viability reduction with values above 50 percent as can be expected from external amine-modified nanosystems [23]. Moreover, the antibiotic-loaded nanosystems showed a statistically significant reduction for all concentrations with respect to unloaded samples, especially the LVX-loaded system that halves the viability of the cells, which could be explained by the initial burst antibiotic released inside cells (**Figure 10**) [60]. This behavior is not isolated, since the decrease in viability and proliferation of osteoblast-like cells when treated with these three antibiotics has been already described in the literature [61-63]. To determine whether this first effect is continued and whether the cells are restored during time, we decided to study the effect of MSNs on the cells after 96 h after treatment (**Figure 11**). In general, we checked that the viability over time was increased even in amine-modified samples as well as in antibiotic-loaded materials, with cell viability values above 60%, demonstrating a rapid recovery of cell viability using these loaded nanosystems.

## 4. Conclusions

This manuscript reports a systematic study of the antimicrobial activity of antibiotic-cargo released from MSNs against different bacterial strains, namely *E. coli* and *S. aureus*. The biological activity of the released doses as well as their effect on bacterial biofilm were evaluated using three different antibiotics. Gentamicin-loaded MSNs exhibited sustained released kinetics without enough antibiofilm effect. However, levofloxacin and rifampin showed release profiles characterizing by an initial burst effect followed by a sustained release with active doses able to reduce up to 99.9 % bacterial biofilm, which remain active for 72 h. The results demonstrate that MSNs are biocompatible and versatile nanocarriers able to load and release diverse antibiotic-cargoes with positive and prolonged antibiofilm effect, which opens up promising expectations to locally treat chronic bone infection. The preliminary in vitro studies to assess the colloidal stability of MSNs in physiological conditions support the formation of a protein corona on the nanoparticles surface. Such protein coverage could compromise the efficiency and biological outcome of the nanoparticles in vivo and therefore much research effort is still needed in the path from bench to bedside.

## Appendices

Information about synthesis and characterization of materials is collect in Supplementary material.

## Acknowledgements

The authors acknowledge the financial support provided by European Research Council (Advanced Grant VERDI; ERC-2015- AdG Proposal No. 694160) and Ministerio de Ciencia e Innovación (Grant MAT2016-75611-R AEI/FEDER).

## Declaration of interest

JE has received funds from Pfizer, Angelini, Biomérieux and Heraeus.

**Tables and Figures**

**Table 1**. Characteristics of the different MSNs synthesized in this work obtained by $N_2$ adsorption porosimetry.

| Sample | $S_{BET}$ (m²/g) | $V_P$ (cm³/g) | $D_P$ (nm) |
|---|---|---|---|
| **MSN** | 1237 | 1.1 | 2.8 |
| **MSN-NH₂** | 786 | 0.5 | 2.3 |
| **MSN-LVX** | 1038 | 0.9 | 2.8 |
| **MSN-NH₂-LVX** | 644 | 0.4 | 2.3 |
| **MSN-GM** | 397 | 0.2 | 2.2 |
| **MSN-NH₂-GM** | 26 | ~ 0 | - |
| **MSN-RIF** | 655 | 0.5 | 2.5 |
| **MSN-NH₂-RIF** | 25 | ~ 0 | - |

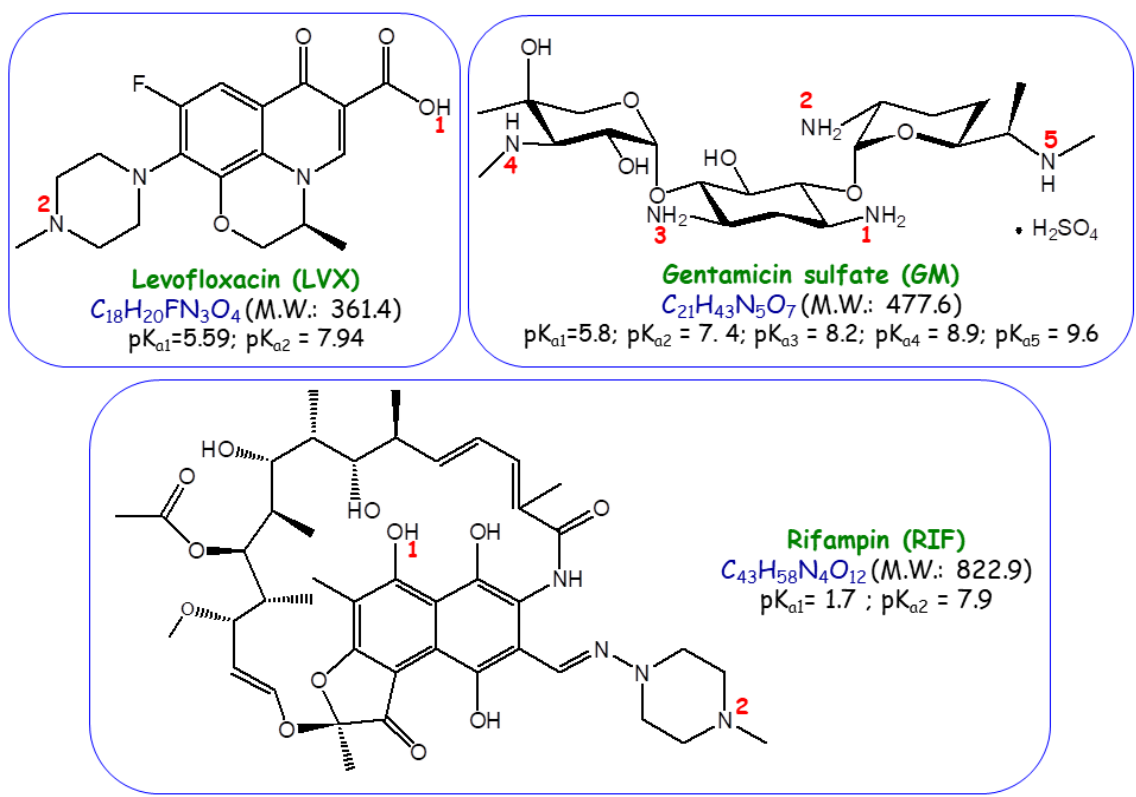

**Figure 1**. Structure and main chemical properties of the three antibiocis used in this work, namely, levofloxacin (LVX), gentamicin (GM) and rifampin (RIF).

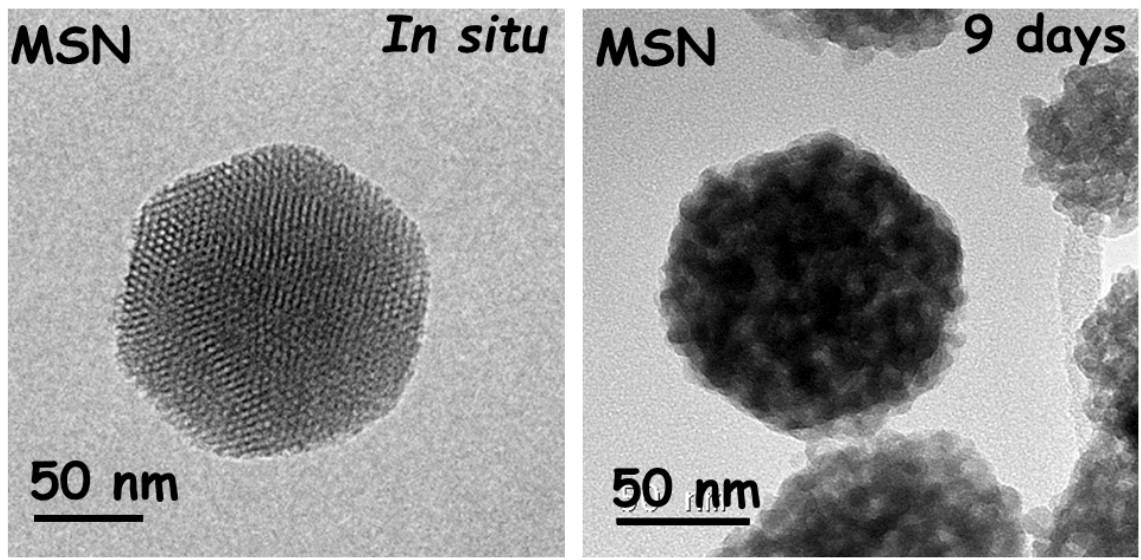

**Figure 2**. TEM images of pristine MSNs before and after 9 days being soaked in PBS under physiological conditions.

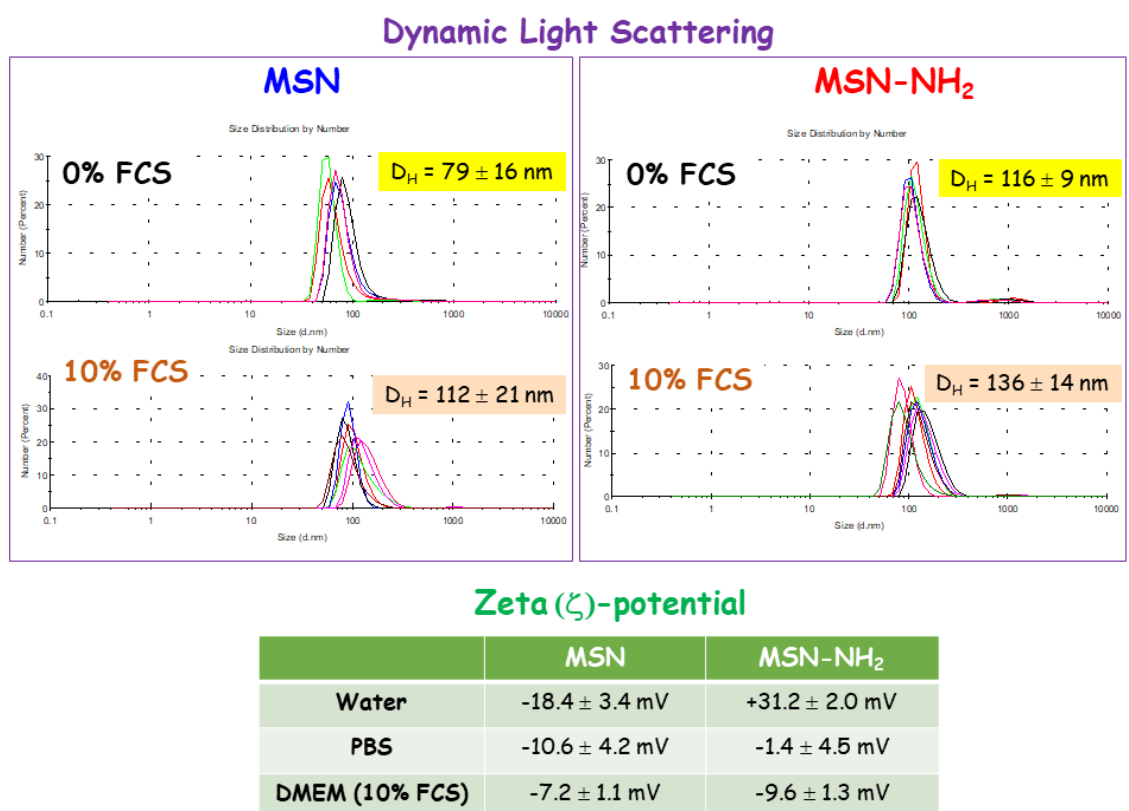

**Figure 3**. Hydrodynamic diameter ($D_H$) by dynamic light scattering (DLS) for MSN and MSN-NH$_2$ after 48 h of incubation with DMEM supplemented with 10% FCS and PBS

1x (0% FCS). Zeta ($\zeta$)-potential values for MSNs in different media are displayed in the inset table.

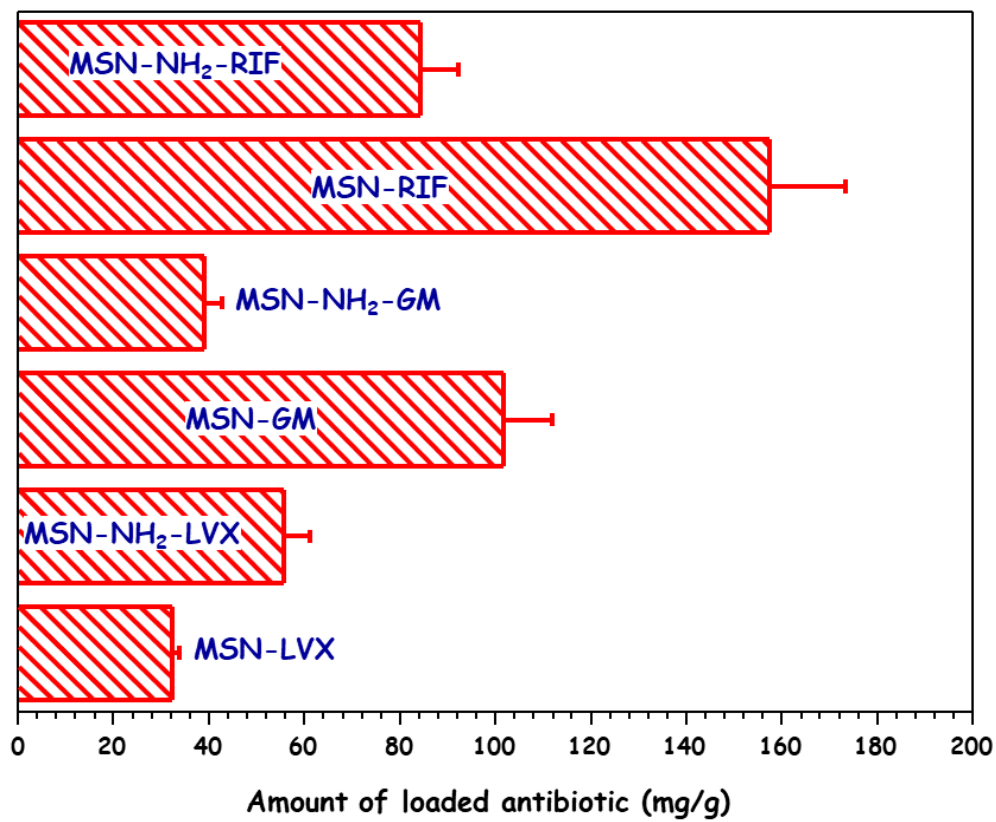

**Figure 4**. Amount of antibiotic loaded into MSN and MSN-NH$_2$ samples determined by elemental chemical analysis.

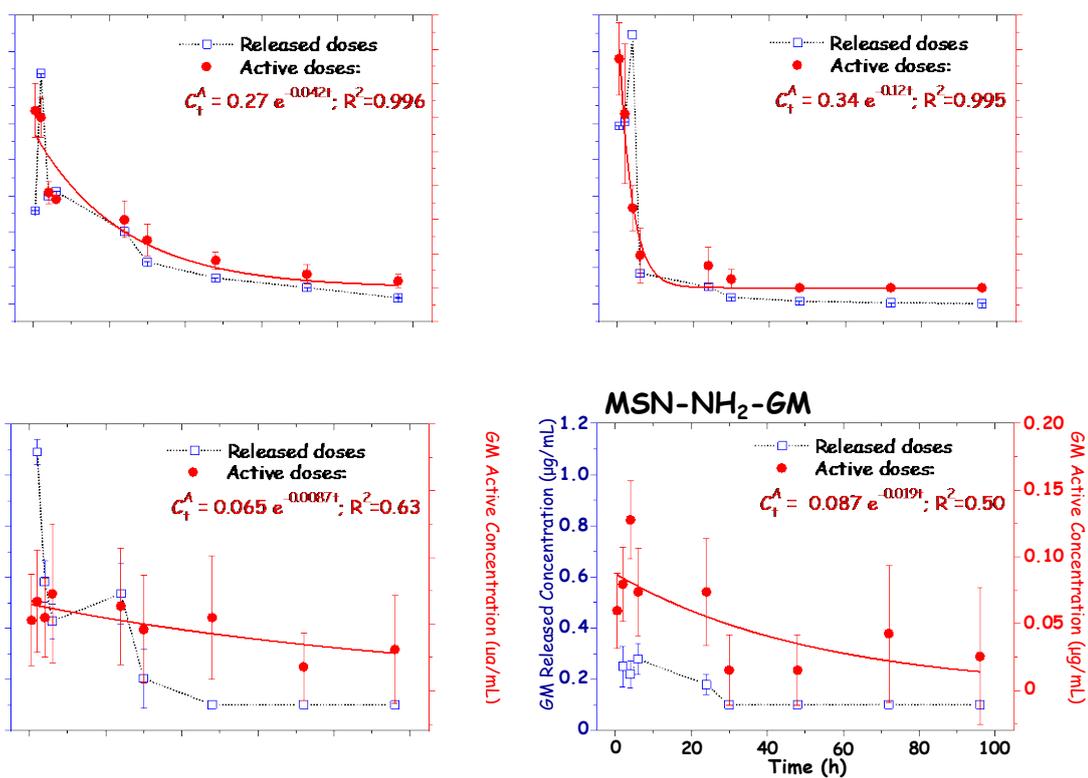

**Figure 5**. Released and active concentrations of antibiotic (LVX and GM) after release from MSN and MSN-NH₂ materials at different time periods determined by disc diffusion tests in planktonic *E. coli* bacteria cultures.

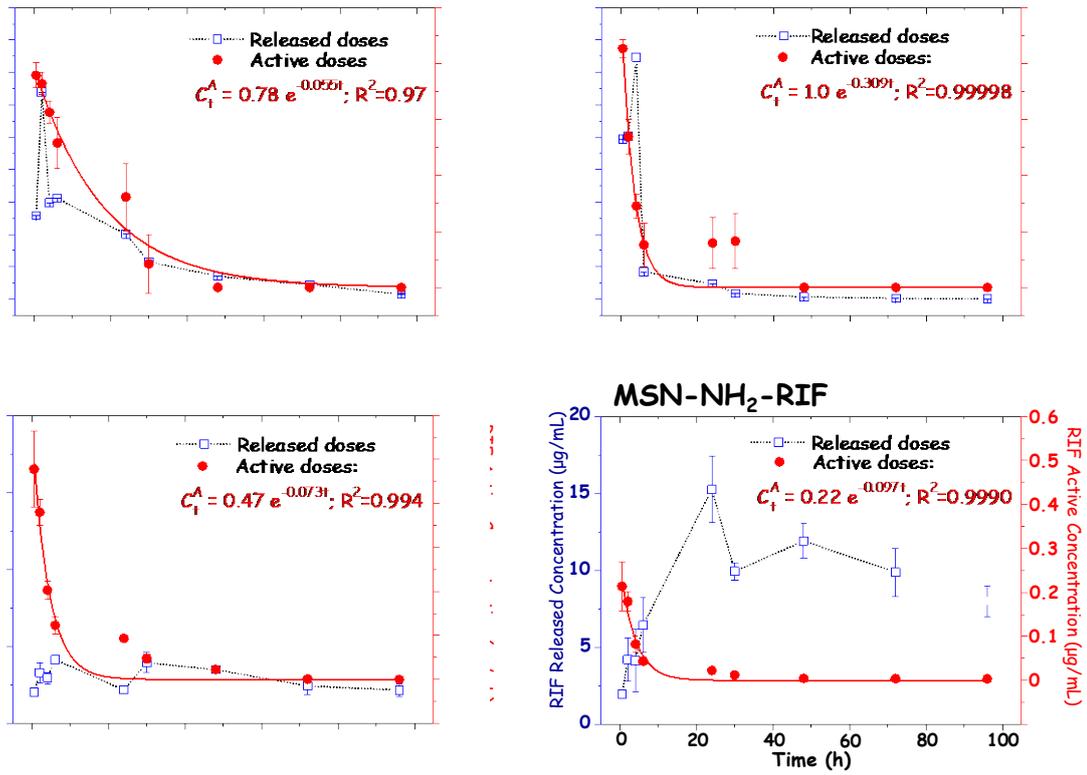

**Figure 6**. Released and active concentrations of antibiotic (LVX and RIF) after release from MSN and MSN-NH₂ materials at different time periods determined by disc diffusion tests in planktonic *S. aureus* bacteria cultures.

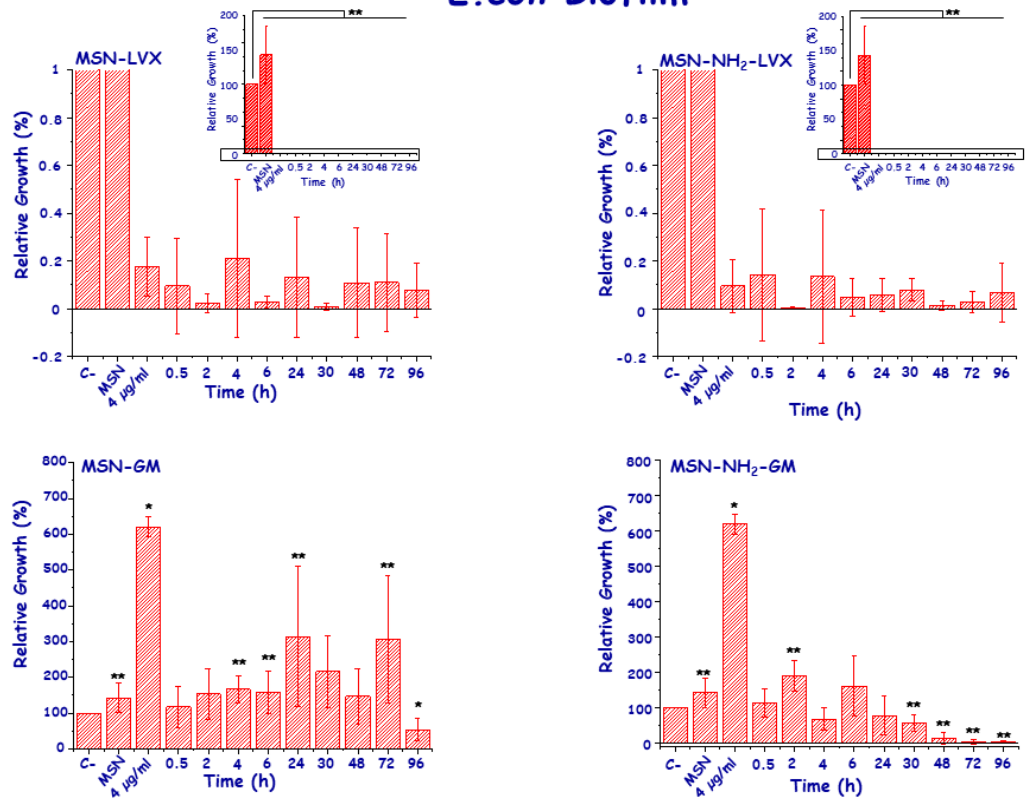

**Figure 7**. *In vitro* antibiofilm activity LVX and GM released from MSN and MSN-NH₂ materials at different times of assay in mature *E. coli* biofilms.

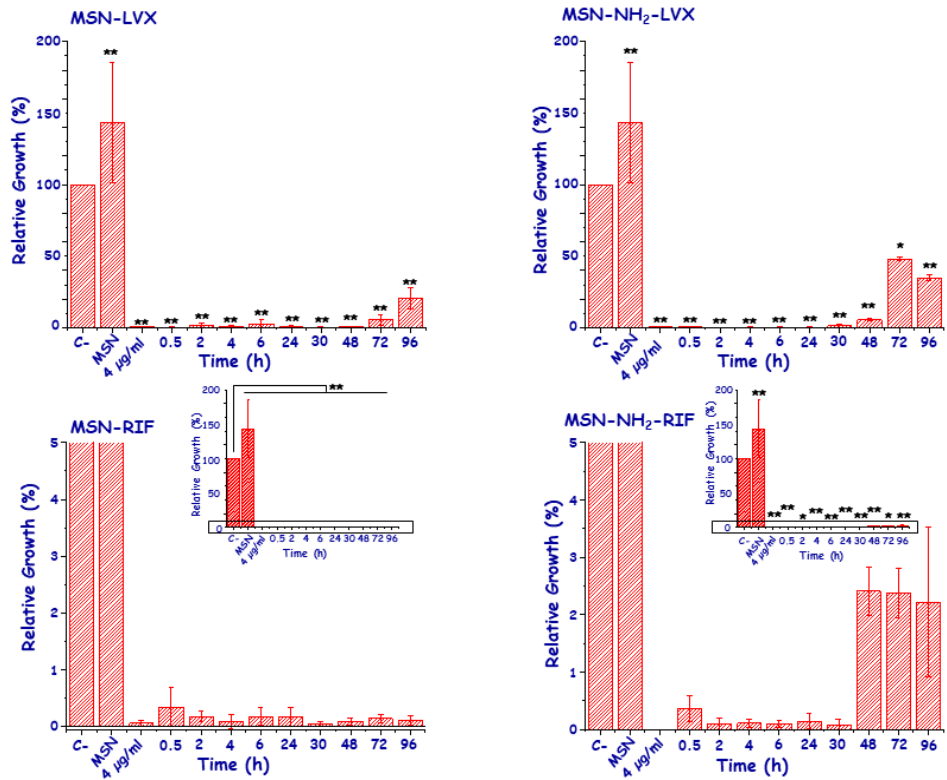

**Figure 8**. *In vitro* antibiofilm activity of LVX and RIF released from MSN and MSN-NH₂ materials at different times of assay in mature *S. aureus* biofilms.

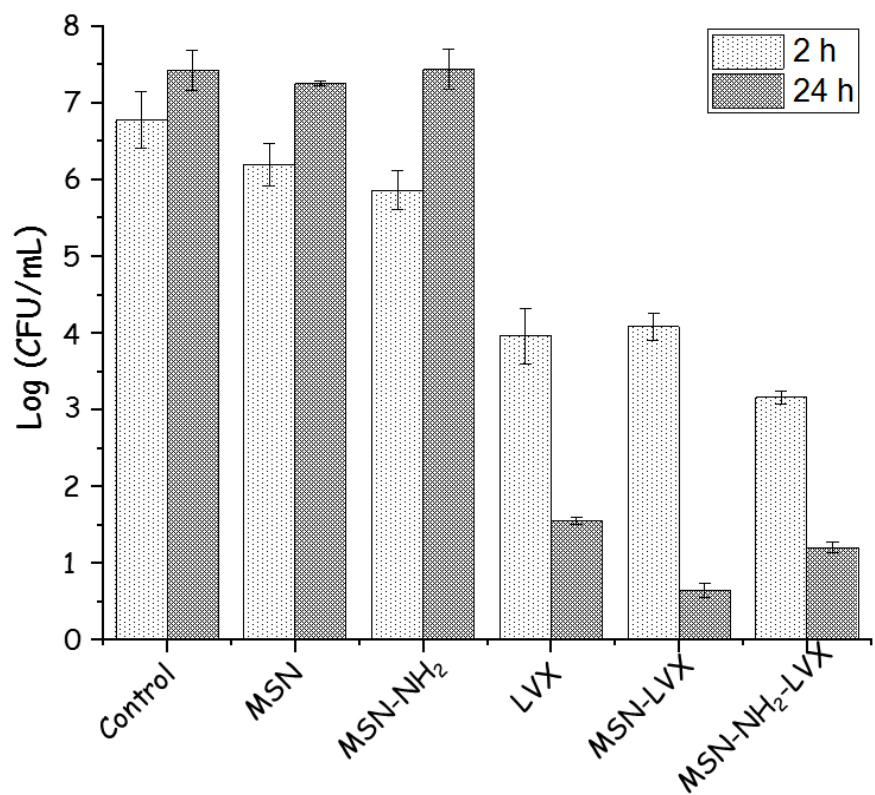

**Figure 9:** Direct effect of the different nanosystems onto *E. coli* biofilm. Histograms represent the Log (CFU/mL) after incubation with the different nanosystems at two tested times (2 and 24 h). Control represents the bacterial control without any treatment. LVX represents the solely antibiotic incubated in the bacteria culture at maximum concentration release by the nanosystems. The effect of antibiotic-free nanosystems on *E. coli* biofilm is also displayed.

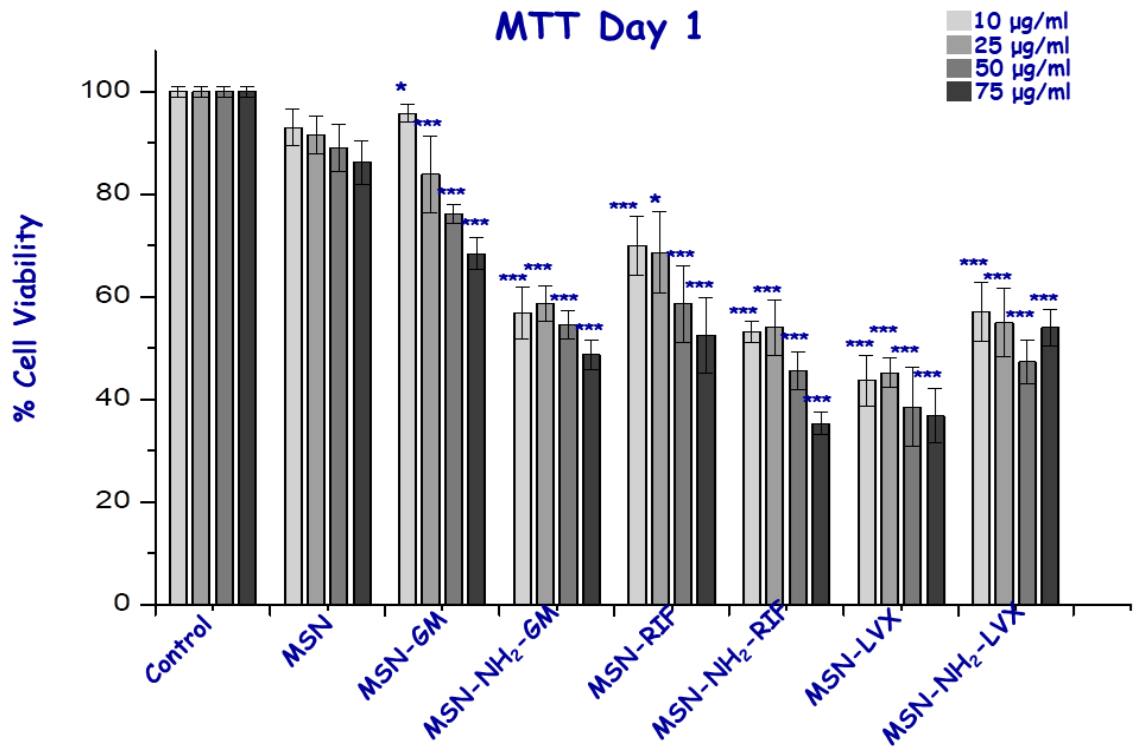

**Figure 10**. Cell viability studies of the samples at different concentrations for the MC3T3-E1 cell line with 1 day of exposure time. *, ** and *** *vs.* corresponding control without nanoparticles (ANOVA).

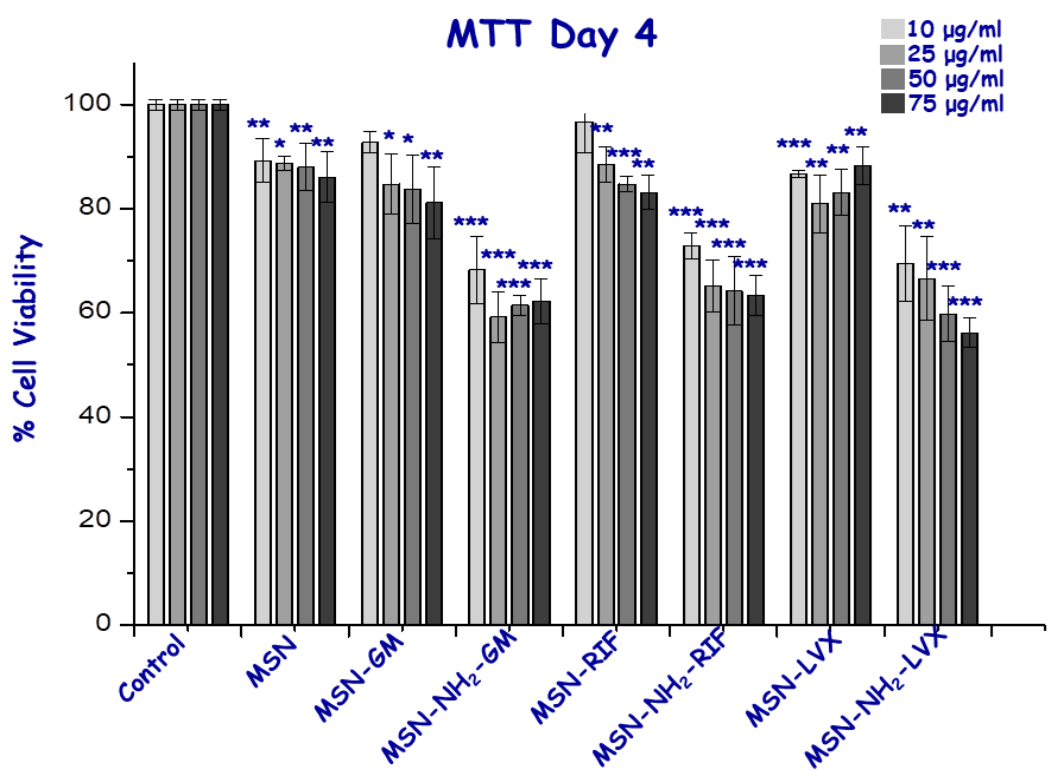

**Figure 11**. Cell viability studies of the samples at different concentrations for the MC3T3-E1 cell line with 4 days of exposure time. *, ** and *** *vs.* corresponding control without nanoparticles (ANOVA).